# Data-driven discovery and rapid, direct synthesis of MXenes

**Authors:** Ali Saffar Shamshirgar[1]*, Guilherme Ribeiro Portugal[1], Soheil Ershadrad[1], Roman Ivanov[2], Martin Dahlqvist[1], Florian Chabanais[3], Sanjay Chakraborty[4], Rainer Traksmaa[2], Irina Hussainova[2], Fredrik Heintz[4], Per O.Å. Persson[2,5], Johanna Rosen[1,5]*

**Affiliations:**

[1]Materials Design, Department of Physics, Chemistry and Biology (IFM), Linköping University; SE-581 83 Linköping, Sweden.

[2]Department of Mechanical and Industrial Engineering, Tallinn University of Technology; 19086 Tallinn, Estonia.

[3]Thin Film Physics, Department of Physics, Chemistry and Biology (IFM), Linköping University; SE-581 83 Linköping, Sweden.

[4]Artificial Intelligence and Integrated Computer Systems, Department of Computer and Information Science (IDA), Linköping University; SE-581 83 Linköping, Sweden.

[5]Wallenberg Initiative Materials Science for Sustainability (WISE), Department of Physics, Chemistry and Biology (IFM), Linköping University, Linköping 58183, Sweden.

*Corresponding author. Email: ali.saffar@liu.se; johanna.rosen@liu.se.

**Abstract:**
MXenes, two-dimensional transition-metal carbides and nitrides, are typically obtained from MAX phases, yet historical reports suggest a broader, largely unexplored chemical space. Here we combine machine-learning–assisted database mining with experiments to uncover overlooked multilayer (ml) MXenes. Screening of repositories reveals a “Treasure Chest” of 38 previously synthesized but unrecognized ml-MXene candidates. Guided by these findings, we rediscover five MXenes using a rapid, scalable self-propagating high-temperature synthesis that requires no sustained external heating and completes within minutes. Inspired by the identified chemistries, we further realize 11 previously unexplored rare-earth-based $M_2CT_2$ MXenes (M= Pr, Nd, Sm, Gd, Tb, Ho, and Tm). Experiments and theory reveal semiconducting behavior and diverse magnetic states across this family. Together, these results expand the MXene family and demonstrate a data-driven strategy for accelerating materials discovery through sustainable methods.

Two-dimensional (2D) transition-metal carbides and nitrides, known as MXenes, have emerged as a rapidly expanding class of versatile materials with exceptional potential for applications in energy storage, electromagnetic interference shielding, catalysis, and superconductivity (*1–5*). First discovered in 2011 (*6*), MXenes are most commonly synthesized by selectively removing the A element from precursor MAX phases using wet-chemical, electrochemical, molten-salt, or gas-phase etching (*6–10*). Direct synthesis routes, such as chemical vapor deposition, has broadened the concept of MXenes beyond MAX-derived structures (*11*). Although this approach typically produces multilayer structures that may subsequently be delaminated (*12*), the resulting MXenes retain unique properties.

Recent advances suggest that MXene-like structures may exist beyond the currently recognized chemical space. Examples include compounds such as $Nb_2CS_2$ (*13*), $Ta_2CSe_2$ (*14*) and $Zr_2CCl_2$ (*15*), which were reported decades before the introduction of MXenes but only recently rediscovered and recognized as belonging to this materials family (*4*, *11*, *16*, *17*). These observations raise a fundamental question: how many MXene-like materials already exist but remain unrecognized within the vast experimental literature and materials databases?

Data-driven research continues to expand rapidly, and methods capable of uncovering hidden information in existing datasets can accelerate materials discovery and inspire new design strategies. Yet the impact of MXenes depends on both discovering new compositions and on how they are synthesized. Most MXenes continue to rely on laboratory-scale processes that involves hazardous reagents (e.g., HF) and multiple time-consuming processing steps (*5*). For MXene research to transition from fundamental discovery to practical technologies, advances in materials design must therefore be accompanied by the development of scalable and sustainable synthesis routes capable of delivering high-quality materials with reduced environmental and energy footprints.

Here, we present a data-driven strategy to uncover historical materials with MXene-like structural motifs. By applying machine learning (ML) to screen large materials databases, we identified a set of previously reported compounds whose structures and chemistries fall within the concept of multilayer (ml) MXenes. We refer to this collection as the *Treasure Chest*, a curated repository of MXene analogues historically classified under different material taxonomies. This framework categorizes MXene-related materials into three groups: (i) compounds originally reported as MXenes, (ii) historical compounds later rediscovered and reclassified as MXenes, and (iii) previously unrecognized MXene-like materials revealed through our ML-assisted search. This approach identified 36 MXene-like candidates synthesized prior to the discovery of MXenes in 2011, and two after. Guided by these findings, we experimentally rediscovered five MXenes from the *Treasure Chest* using a sustainable and scalable direct synthesis route, previously not applied to MXenes, enabling the formation of the target phases within minutes. Inspired by the expanded chemical space, we further synthesized 11 previously unexplored MXenes. Importantly, while structural data for $Er_2CCl_2$ MXene appear in a repository (*18*), our work expands the MXene family by adding seven rare-earth (RE) elements (Pr, Nd, Sm, Gd, Tb, Ho, and Tm). For the 17 experimentally obtained phases, we performed a systematic investigation of structural, electronic, optical, and magnetic properties using complementary experimental and theoretical approaches. In addition, 26 hypothetical phases are predicted to be stable and were investigated theoretically.

Our findings demonstrate a substantial reservoir of overlooked MXene-like materials already documented in the experimental literature and available materials databases. Beyond identifying specific compounds, this work establishes a general and scalable strategy for the inverse discovery of functional materials hidden in existing data. By systematically mining historical records for unrecognized structural motifs and overlooked compounds, we provide a roadmap for accelerating the discovery of not only 2D materials but a broad range of material classes relevant to present technological challenges.

## Results

We set out to expand the pool of experimentally accessible MXene phases beyond those currently recognized under the "MXene" label, including compounds reported decades ago in other contexts. To do this systematically, we screened two large crystal-structure repositories, the Materials Project (MP) and the Open Quantum Materials Database (OQMD) (*19–21*). Here, we use the term multilayered MXene (ml-MXene) to denote ternary layered materials built from $M_2X$ slabs (M = transition metal, lanthanide, actinide, X = C, N or related light elements) where a third element (T) occupies surface termination sites on the $M_2X$ slabs, such that delamination would yield two-dimensional $M_2XT_2$ sheets (Fig. 1A).

### *Data-driven search for ml-MXenes*

The screening was restricted to ternary compounds with a 2:1:2 stoichiometry consistent with an $M_2XT_2$ phase (allowing any permutation of the three elements in the reduced formula). We constrained the element pool to compositions chemically adjacent to known ml-MXene phases, assigning M from a broad set of transition and rare-earth (RE) metals (Sc, Y, Ti, Zr, Hf, V, Nb, Ta, Cr, Mo, W, Mn, Co, Bi, Sb, La, Ce, Pr, Nd, Pm, Sm, Eu, Gd, Tb, Dy, Ho, Er, Tm, Yb, Lu, Pu, Th, U), X from light elements (C, N, O, P, S, B), and T from common termination species (O, S, Se, Te, N, F, Cl, Br, I). After merging the outputs from the two databases, a total of 541 candidates were identified and their CIF files retrieved. (Table S1). A CIF-based classifier was then developed to identify candidates with ml-MXene structural motifs. We benchmarked four supervised classifiers: support vector machine, multilayer perceptron, a simplified graph-based model, and Random Forest (Fig. S1-3). Training was performed on a balanced dataset of 1050 labeled structures comprising 525 ml-MXene and 525 non-ml-MXene structures (Table S1). Within the ml-MXene subset, the most frequent space groups were $P6_3/mmc, P\bar{3}m1$, and $R\bar{3}m$ (Fig. 1B), whereas the non-ml-MXene subset was intentionally heterogeneous and spanned a broader range of space groups. For each structure, the workflow parses the CIF file and computes structural descriptors including global symmetry information, pair-distance statistics, density-related features, and measures of structural heterogeneity. The Random Forest model achieved the best overall performance, and its predicted class probability was therefore used to triage the candidate list (Table S2). Using a decision threshold of $\mathcal{P}$(ml-MXene) > 0.8, we filtered the candidate list, leaving 270 materials for further analysis (Table S3).

Since the focus was on compositions with experimental precedents, the candidate list was filtered to retain database entries with experimental tags, reducing the set to 33 materials. However, database annotations evolve over time and literature tags often lag behind. To keep the present methodology generic, while still comprehensive and practical, we complemented the metadata filter with targeted literature searches for additional ml-MXene candidates. This step expanded the set to include 17 additional compositions that were labeled as ml-MXene candidates by our classifier but had a missing experimental tag in MP or OQMD, despite independent evidence of synthesis: $Hf_2PTe_2$, $Mo_2CCl_2$, $Ti_2CCl_2$, $Ti_2CBr_2$, $Ti_2CTe_2$, $Ti_2NCl_2$, $Ta_2CSe_2$, $Nb_2CSe_2$, $Nb_2CCl_2$, $Gd_2CBr_2$, $Y_2CF_2$, $Y_2CCl_2$, $Y_2CBr_2$, $Y_2PBr_2$, $Zr_2CBr_2$, $Zr_2CCl_2$, and $Zr_2NCl_2$. Restricting this final step to fully terminated phases ($T_2$) yielded a curated set of 50 materials. An overview of the screening, classification, and selection workflow is shown in Fig. 1C.

For clarity, the 50 materials were grouped into three categories (Table 1). Group I includes 8 phases synthesized after the first MXene report in 2011 (*6*). This set is dominated by titanium- and niobium-based phases, and all eight entries, including $Y_2CF_2$ and $Mo_2CCl_2$, were described as MXenes in their respective references. Group II comprises $Nb_2CS_2$, $Ta_2CS_2$, $Ta_2CSe_2$, and $Zr_2CCl_2$, which were synthesized before the introduction of MXenes (2011), but later rediscovered and reinterpreted within the MXene framework. For example, multilayered niobium carbosulfide was first obtained through topochemical reactions in 1992 (*13*) and later accessed through covalent surface modification routes applied to pre-existing MXenes in 2020 (*4*). Similarly, chlorine-terminated zirconium carbide was initially synthesized through high-temperature reactions of stoichiometric precursors (750-1000 °C) (*15*) and has more recently been prepared by chemical vapor deposition (*11*). Group II illustrates how compositions known from earlier solid-state chemistry can be revisited using modern synthesis approaches with improved compositional and structural control. The remaining 38 compositions were experimentally reported, often decades prior to the emergence of MXenes, but have not been recognized or investigated as such. We refer to Group III as the "*Treasure Chest*", as it comprises multilayered phases that remain largely unexplored despite clear structural correspondence with MXenes (Fig. 1).

The *Treasure Chest* demonstrates that the experimental literature already contains a substantial set of ml-MXene-like phases that have not yet been explored within the MXene context. These compounds span 20 distinct M-site elements, 16 of which (Ce, Dy, Er, Eu, Gd, Ho, La, Lu, Nd, Pr, Sm, Tb, Tm, Yb, Th, and U) are not represented among experimentally reported MXenes (*10*). This set also extends the multilayer chemistry beyond the carbon/nitrogen focus of established MXenes to also include X-site elements O, S, and P. Consequently, the *Treasure Chest* provides experimentally grounded starting points for exploring new ml-MXene chemistries outside the conventional transition-metal carbide and nitride space. For target selection in the present work, we used a compact site-resolved summary over the screened dataset (Fig. S4) to identify combinations most consistently classified as ml-MXenes by the model. This analysis identified RE–containing carbides with halogen terminations as the most promising region within the screened phase space, guiding re-synthesis and first-synthesis experimental efforts.

**Table 1.** Multilayer-MXenes identified by screening the MP and OQMD databases, prioritized by the CIF-based classifier. For each phase, we list the M, X, and T components, the calculated model probability of being a ml-MXene structure, $\mathcal{P}$(ml-MXene), and the year of the earliest experimental report. The 50 phases were grouped as Group I ("Discovered MXenes"), Group II ("Rediscovered MXenes"), and Group III ("*Treasure Chest*").

| | Phase | M | X | T | $\mathcal{P}$ (ml-MXene) | Year | Ref. |
|---|---|---|---|---|---|---|---|
| **Group I Discovered MXenes** | $Nb_2CSe_2$ | Nb | C | Se | 0.982 | 2020 | (*4*) |
| | $Nb_2CCl_2$ | Nb | C | Se | 0.974 | 2020 | (*4*) |
| | $Ti_2CCl_2$ | Ti | C | Cl | 0.990 | 2020 | (*22*) |
| | $Ti_2CBr_2$ | Ti | C | Br | 0.968 | 2020 | (*4*) |
| | $Ti_2CTe_2$ | Ti | C | Te | 0.978 | 2020 | (*4*) |
| | $Y_2CF_2$ | Y | C | F | 0.860 | 2019 | (*23*) |
| | $Mo_2CCl_2$ | Mo | C | Cl | 0.970 | 2026 | (*24*) |
| | $Ti_2NCl_2$ | Ti | N | Cl | 0.990 | 2020 | (*11*) |
| Group II Rediscovered MXenes | $Nb_2CS_2$ | Nb | C | S | 0.978 | 1992<br>2020 | (*13*)<br>(*4*, *16*) |
| | $Ta_2CS_2$ | Ta | C | S | 0.986 | 1970<br>2022 | (*11*)<br>(*16*) |
| | $Ta_2CSe_2$ | Ta | C | Se | 0.986 | 1993<br>2025 | (*14*)<br>(*17*) |
| | $Zr_2CCl_2$ | Zr | C | Cl | 0.980 | 1986<br>2023 | (*15*)<br>(*11*) |
| Group III " Treasure Chest" | $Gd_2CBr_2$ | Gd | C | Br | 0.974 | 1985 | (*25*) |
| | $Gd_2CCl_2$ | Gd | C | Cl | 0.966 | 1994 | (*26*) |
| | $Gd_2CF_2$ | Gd | C | F | 0.874 | 1991 | (*27*) |
| | $Ho_2CF_2$ | Ho | C | F | 0.848 | 1992 | (*28*) |
| | $Lu_2CCl_2$ | Lu | C | Cl | 0.952 | 1987 | (*29*) |
| | $Sc_2CCl_2$ | Sc | C | Cl | 0.802 | 1986 | (*15*) |
| | $Y_2CCl_2$ | Y | C | Cl | 0.958 | 1986 | (*15*) |
| | $Y_2CBr_2$ | Y | C | Br | 0.974 | 1995 | (*30*) |
| | $Zr_2CBr_2$ | Zr | C | Br | 0.976 | 1986 | (*15*) |
| | $Sc_2NCl_2$ | Sc | N | Cl | 0.984 | 1986 | (*15*) |
| | $Zr_2NCl_2$ | Zr | N | Cl | 0.988 | 1986 | (*15*) |
| | $Th_2ON_2$ | Th | O | N | 0.906 | 1966 | (*31*) |
| | $Ce_2SO_2$ | Ce | S | O | 0.988 | 1949 | (*32*) |
| | $Dy_2SO_2$ | Dy | S | O | 0.954 | 1958 | (*33*) |
| | $Er_2SO_2$ | Er | S | O | 0.968 | 1958 | (*33*) |
| | $Eu_2SO_2$ | Eu | S | O | 0.914 | 1958 | (*33*) |
| | $Gd_2SO_2$ | Gd | S | O | 0.972 | 1958 | (*33*) |
| | $Ho_2SO_2$ | Ho | S | O | 0.942 | 1958 | (*33*) |
| | $La_2PBr_2$ | La | P | Br | 0.800 | 2007 | (*34*) |
| | $La_2PI_2$ | La | P | I | 0.800 | 2007 | (*34*) |
| | $La_2SO_2$ | La | S | O | 0.948 | 1949 | (*32*) |
| | $Lu_2SO_2$ | Lu | S | O | 0.960 | 1958 | (*33*) |
| | $Nd_2SO_2$ | Nd | S | O | 0.956 | 1958 | (*33*) |
| | $Pr_2SO_2$ | Pr | S | O | 0.976 | 1958 | (*33*) |
| | $Sc_2SO_2$ | Sc | S | O | 0.888 | 1978 | (*35*) |
| | $Sm_2SO_2$ | Sm | S | O | 0.972 | 1958 | (*33*) |
| | $Tb_2SO_2$ | Tb | S | O | 0.978 | 1958 | (*33*) |
| | $Th_2SN_2$ | Th | S | N | 0.956 | 1969 | (*36*) |
| | $Tm_2SO_2$ | Tm | S | O | 0.962 | 1958 | (*33*) |
| | $U_2PN_2$ | U | P | N | 0.952 | 1969 | (*36*) |
| | $U_2SN_2$ | U | S | N | 0.960 | 1969 | (*36*) |
| | $Y_2PBr_2$ | Y | P | Br | 0.890 | 2007 | (*34*) |
| | $Y_2SO_2$ | Y | S | O | 0.958 | 1969 | (*37*) |
| | $Yb_2SO_2$ | Yb | S | O | 0.916 | 1958 | (*33*) |
| | $Zr_2SN_2$ | Zr | S | N | 0.912 | 2003 | (*37*) |
| | $Zr_2PTe_2$ | Zr | P | Te | 0.816 | 2009 | (*38*) |
| | $Hf_2PTe_2$ | Hf | P | Te | 0.814 | 2016 | (*39*) |
| | $Hf_2SN_2$ | Hf | S | N | 0.910 | 2013 | (*40*) |

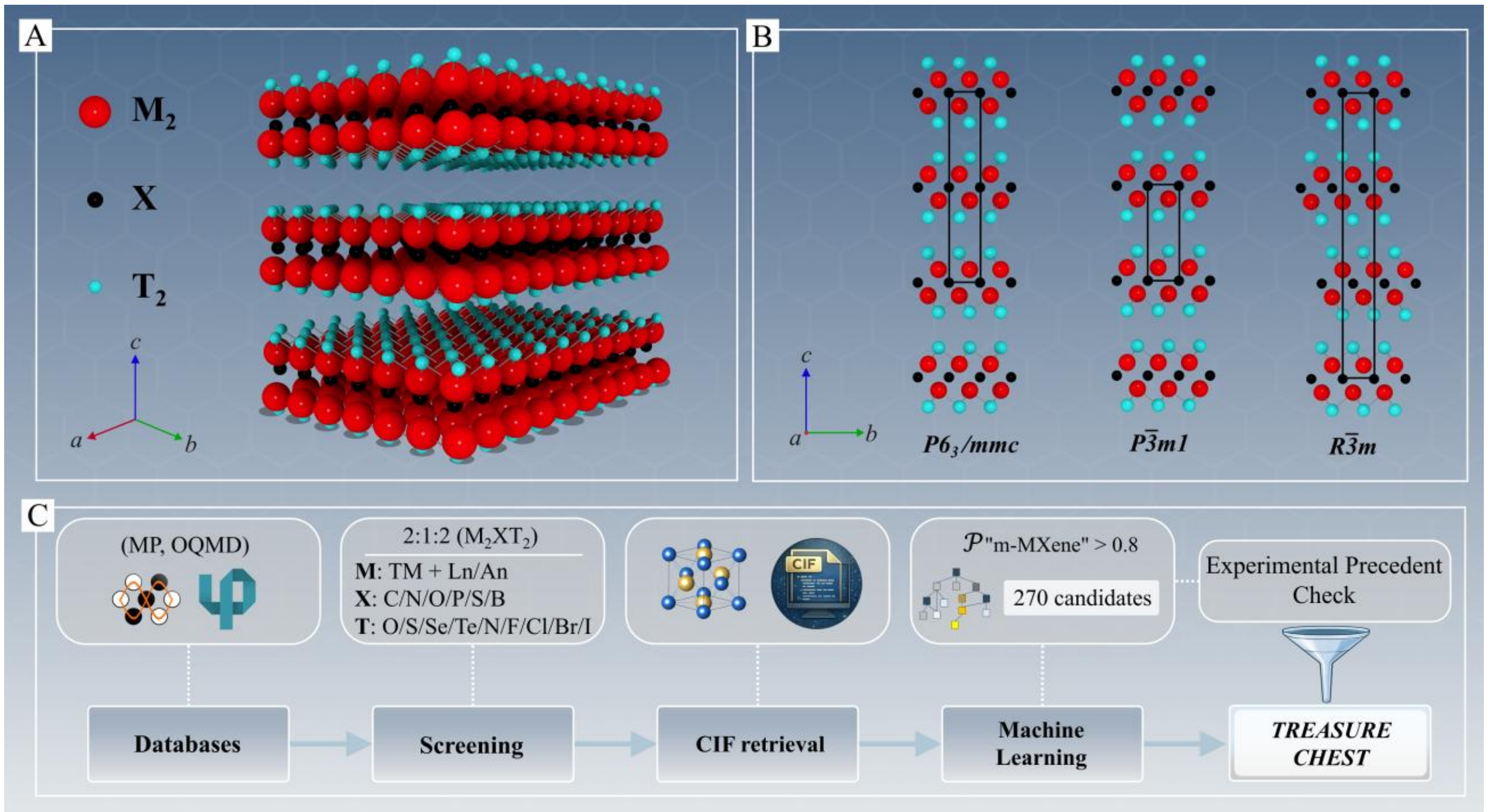

**Fig 1.** Data-driven discovery framework for ml-MXene compositions. (A) Perspective view of a multilayered $M_2XT_2$ MXene structure. (B) Representative crystal structures of the most frequently observed space groups for ml-MXene structures ($P6_3/mmc, P\bar{3}m1$ and $R\bar{3}m$). (C) Workflow used to identify potential ml-MXene candidates: crystal structures were mined from MP and OQMD databases, filtered to reduced 2:1:2 stoichiometry using predefined element pools, retrieved as CIF files, and classified with a machine-learning model. A total of 270 candidates satisfied $\mathcal{P}$(ml-MXene) > 0.8 and were subsequently evaluated for experimental precedent, yielding 50 experimentally reported compositions, of which 38 were previously synthesized but not recognized as ml-MXene phases (*Treasure Chest*).

*Direct synthesis of ml-MXenes*

Inspired by the *Treasure Chest* and by earlier reports of direct synthesis routes for related materials, we sought to explore previously unreported MXenes while establishing alternative synthesis approaches that are rapid, scalable, and sustainable. We selected $Y_2CF_2$ (known MXene), $Y_2CBr_2$, and $Y_2CCl_2$ as representative systems to demonstrate the feasibility of direct self-propagating high-temperature synthesis (SHS). In this approach, stoichiometric yttrium is reacted with polytetrafluoroethylene (PTFE, $(C_2F_4)_n$), whose feasibility relies on the self-sustaining nature of the SHS reaction. To assess this, phase stability and the adiabatic reaction temperature ($T_{ad}$) were evaluated following the procedure described in (*41*). The calculated adiabatic temperatures are 2228 K for the direct elemental reaction (mixing the constituent elements) and 1627 K for the exothermic SHS reaction $4M + (C_2F_4)_n = 2M_2CF_2$ (Table S4). The experimental combustion temperature was measured at 1740 K.

As illustrated in Fig. 2A, ignition by brief Joule heating of a tungsten coil (~5 s) locally raises the temperature, initiating depolymerization of $(C_2F_4)_n$ at 300-400 °C to form $C_2F_4$ monomers (*42*). Subsequent thermal decomposition at ~660 °C generates $CF_x$ (x = 1–3) radical species (*41*, *43*), which act as the reactive fluorine source for $Y_2CF_2$ formation. Once initiated, the reaction proceeds in a self-sustaining manner without external heating until the precursors are consumed, completing within ~5 min including cooling to room temperature. Because yttrium strongly favors a +3 oxidation state, excess fluorine released during the reaction can partially fluorinate metallic Y to form $YF_3$ (*44*). Introducing excess carbon relative to the stoichiometric composition suppresses this competing reaction and promotes $Y_2C$ formation, increasing the phase purity of $Y_2CF_2$ from ~60% to ~90% as confirmed by X-ray diffraction (XRD) Rietveld refinement (Fig. 2B) (Table S5). The resulting product is obtained as a uniform green pellet that remains structurally intact after synthesis. Additional details of the parametric study are provided in Fig. S5-6.

A similar strategy was applied to synthesize $Y_2CBr_2$ using carbon tetrabromide ($CBr_4$) as the bromine precursor according to the reaction $4M + CBr_4 + C = 2M_2CBr_2$ (Fig. 2D). Carbon tetrabromide melts at 367 K and decomposes at ~463 K (*45*), leading to rapid and strongly exothermic reactions when used as a brominating agent. The calculated $T_{ad}$ for $Y_2CBr_2$ formation is 2519 K, with an experimentally measured combustion temperature of 2362 K for synthesis using a compacted pellet. To mitigate uncontrolled bromine outgassing and moderate the reaction kinetics, bromine-terminated samples were instead synthesized in powder form using a graphite-lined alumina crucible with a fitted lid. This modification increases the phase purity from ~45% in the pressed-pellet geometry to >85% when the reaction was performed using loose powder. The corresponding Rietveld refinement is shown in Fig. 2E. Additional experimental details and Rietveld refinement data are provided in Fig. S6, and Table S6.

An analogous strategy was used to synthesize $Y_2CCl_2$ using Hexachloroethane ($C_2Cl_6$) as the chlorine precursor according to the reaction $6M + C_2Cl_6 + C = 3M_2CCl_2$. Hexachloroethane sublimes at ~187 °C, initiating decomposition into chlorinated intermediates such as $CCl_4$ and $C_2Cl_4$, followed by further chain pyrolytic decomposition reactions leaving C and releasing $Cl_2$ (*46*). The resulting $Y_2CCl_2$ exhibits a brown color, high phase purity, and distinct layered structure (Fig. S7).

Scanning transmission electron microscopy (STEM) images of $Y_2CF_2$ acquired along the $[0001]$ and $[11\bar{2}0]$ zone axes (Fig. 2 H,J,K) and Rietveld refinement (Fig. 2B) confirms that $Y_2CF_2$ crystallizes in the trigonal $P\bar{3}m1$ space group (No. 164), consistent with the selected-area electron diffraction (SAED) pattern in Fig. 2G and the structural overlay of the [0001] projection in Fig. 2H. Moreover, the $[11\bar{2}0]$ projection (Fig. 2 J-K) reveals two Y layers stacked along the $c$-axis, with $c$ = 6.29 Å obtained from Rietveld refinement. For comparison, the bromine-terminated analogue $Y_2CBr_2$ is shown in Fig. 2M-N. The $[11\bar{2}0]$ projection along with the inset in Fig. 2M shows a side-view STEM image with structural overlay, highlighting a single $Y_2C$ slab symmetrically terminated by clearly visible Br atoms. Compared with the F-terminated system, the larger Br termination results in increased interlayer separation, consistent with the larger lattice parameters determined from Rietveld refinement ($a = b$ = 3.74 Å and $c$ = 9.75 Å).

Electronic band gaps, infrared (IR), and Raman spectra were systematically calculated for all systems to facilitate identification of the target compounds. The experimentally determined band gap of $Y_2CF_2$ (2.09 eV; Fig. 2O-P) is in good agreement with the theoretical prediction (1.96 eV) and slightly higher than the previously reported value (1.9 eV) (*23*). This modest deviation may arise from minor phase impurities (5–10%) or defects, as well as uncertainties associated with the Kubelka-Munk-based Tauc extrapolation used to estimate the band gap. First-principles vibrational analysis predicts several Raman-active modes (Fig. 2Q). The most intense mode occurs at ~245 cm$^{-1}$ and corresponds to an $A_{1g}$ vibration, followed by a second prominent $A_{1g}$ mode at ~367 cm$^{-1}$. Two nearly degenerate modes at ~320 cm$^{-1}$ and ~138 cm$^{-1}$ are assigned to the doubly degenerate $E_g$ representation. The experimental Raman spectrum (Fig. 2R) resolves all predicted modes, as confirmed by the deconvolution analysis shown in the inset. The $E_g$ peak associated with in-plane Y-C vibrations closely matches the calculated frequency, whereas the $A_{1g}$ mode appears at 254 $cm^{-1}$, corresponding to a blue shift of ~9 $cm^{-1}$ relative to theory. This mode is assigned to out-of-plane M-T (Y-F) stretching vibration (Fig. 2O, inset). The blue shift is consistent with the slightly smaller experimental lattice parameter ($c$ = 6.29 Å) compared with the DFT value ($c$ = 6.35 Å), which increases the effective force constant of the Y–F bond. Bulk compositional analysis and TEM/EDX measurements further confirm a 1:1 M:T ratio for both F (Fig. S5) and Br (Fig. S6) terminated systems. Additional theoretical and experimental characterization of $Y_2CF_2$, $Y_2CBr_2$, and $Y_2CCl_2$ is provided in Fig. S5-7 and Table S5-7.

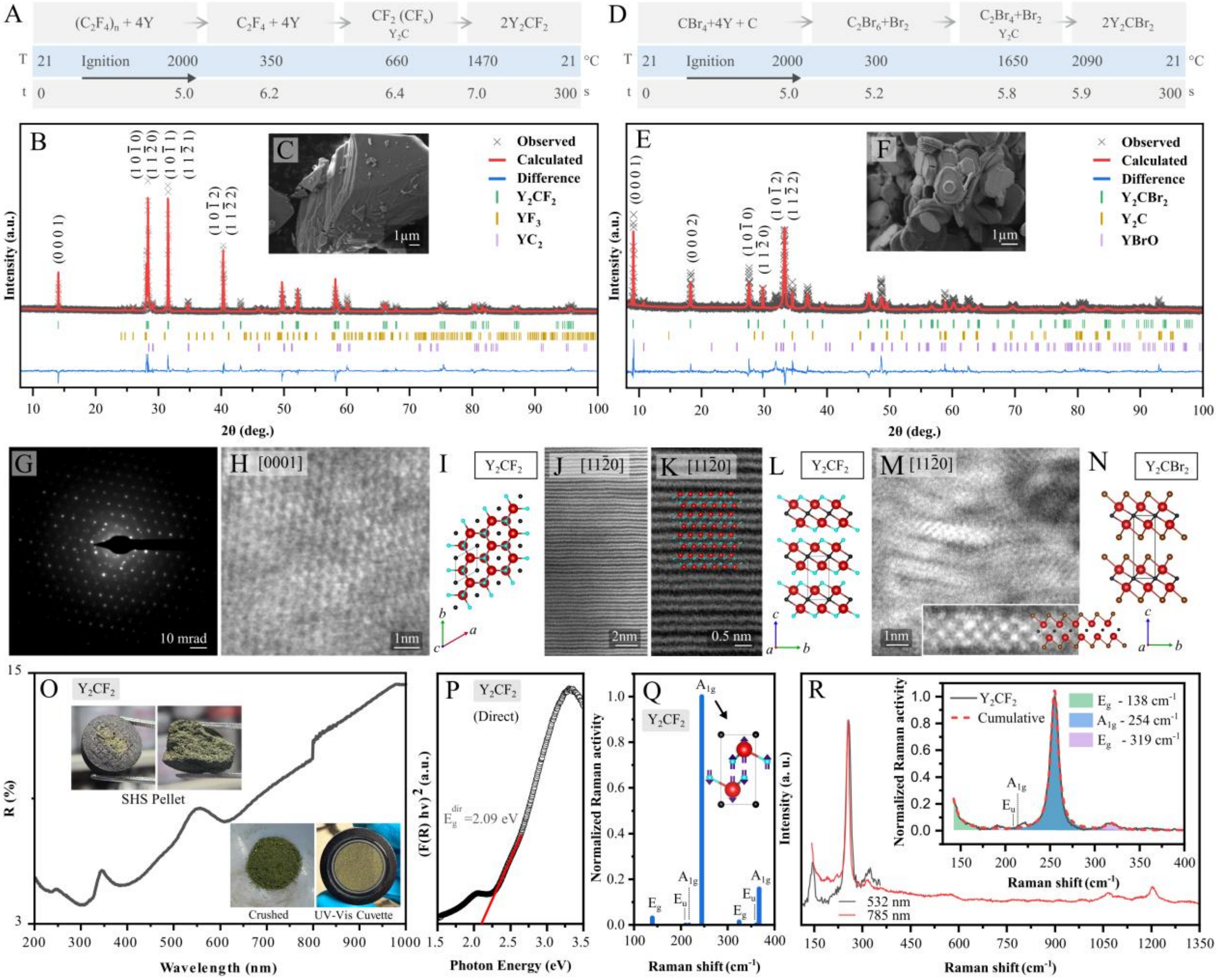

**Fig. 2.** Direct synthesis and characterization of SHS-derived MXenes. Schematic of the SHS reaction pathway with respect to chemistry, temperature, and time, Rietveld refinement, and SEM images of the synthesized powders for (A-C) $Y_2CF_2$ and (D-F) $Y_2CBr_2$. (G) Selected-area electron diffraction (SAED) pattern of SHS-synthesized $Y_2CF_2$ acquired along the [0001] zone axis. (H) Corresponding STEM image along [0001](top view). (I) Schematic top-view projection of the crystal structure. (J, K) STEM images of $Y_2CF_2$ recorded along the $[11\bar{2}0]$ zone axis (side view). (L) Schematic side-view projection of the crystal structure. (M) STEM image of $Y_2CBr_2$ along $[11\bar{2}0]$ (side view) with inset showing side-view of a MXene layer. (N) Schematic side-view projection of the $Y_2CBr_2$ crystal structure. (O) UV-Vis diffuse reflectance spectrum of SHS-synthesized $Y_2CF_2$, with inset photographs of the pellet before and after crushing. (P) Tauc plot

derived from the Kubelka-Munk function. (Q) Calculated Raman spectrum (inset: eigenvectors of the $A_{1g}$ mode shown with a 2x scale factor). (R) Experimental Raman spectrum of $Y_2CF_2$ with inset showing peak deconvolution.

Inspired by the *Treasure Chest* and the expanded MXenes chemical space it revealed, we next explored previously unreported MXene compositions. Following the successful synthesis of ml-MXenes $Y_2CF_2$ and $Y_2CBr_2$ (Figure 2), as well as $Y_2CCl_2$, we investigate $M_2CT_2$ phases with M = Pr, Nd, Sm, Gd, Tb, Ho, and Tm and T = F or Br. Figure 3A provides an overview of all 17 synthesized phases, including $Y_2CF_2$, $Y_2CBr_2$, and $Y_2CCl_2$, with photographs of the powders stored in argon shown beneath the corresponding theoretically calculated band gaps. 11 of these phases (orange circles) are reported here for the first time. The inset shows photographs of the synthesized powders after 24 h of air exposure. SEM, EDX, XRD, UV-Vis, and Raman spectroscopy data for all systems are provided in Fig. S8-21 and Table S8-10. A detailed description and time-resolved visual evolution of the samples during air exposure for up to one month are provided in Fig. S22.

Representative results from synthesis and characterization of $Tm_2CF_2$, $Ho_2CBr_2$, and $Pr_2CBr_2$ are shown in Figure 3. For $Tm_2CF_2$, Rietveld refinement (Fig. 3B) indicates a ~54% phase purity, with impurity phases identified as $Tm_2C_3$ and $TmF_3$. Strategies for improving phase purity are discussed in the Supplementary Information. EDX analysis of the layered microstructure (Fig. 3C) confirms a stoichiometric Tm:F ratio of approximately 1:1 with minimal oxygen content (Fig. 3D). Prior to analysis, the sample was stored under argon and exposed to ambient air for ~10 s during transfer to the SEM chamber. The low oxygen signal indicates that the bulk composition is largely preserved, with any oxidation limited to the surface region.

The UV-Vis spectrum of $Tm_2CF_2$ (Fig. 3E) and the corresponding Kubelka–Munk-derived Tauc plot (Fig. 3F) shows an optical band gap of 2.07 eV, in close agreement with the theoretical prediction of 2.02 eV. As a general note, across the 45 $M_2CT_2$ compositions investigated (M=Y and RE elements; T=F, Br, and Cl), all phases are predicted to be semiconducting, except for the semi-metallic $Ce_2CCl_2$ and $Ce_2CBr_2$. Optical photographs illustrate synthesized $Tm_2CF_2$ powder before and after crushing (Fig. 3G) and upon air exposure from several seconds to one week (Fig. 3H). The freshly prepared sample exhibits a green color that gradually changes to light red upon ambient exposure, accompanied by an apparent increase in volume, reflecting oxidation or hydrolysis. Despite that, XRD shows the presence of MXene after one month of exposure to ambient as evident in Fig S10 (see Supplementary Information for details). The calculated Raman spectrum (Fig. 3I), and the deconvolution of the experimental spectrum (Fig. 3J) shows peak positions closely matching the calculated vibrational frequencies. Further, the *c*-lattice parameter obtained from theory (6.22 Å) closely matches the value determined experimentally from Rietveld refinement. This agreement indicates minimal lattice distortion between the modeled and synthesized structures and explains the absence of a measurable shift in the $A_{1g}$ Raman mode.

STEM images of $Tm_2CF_2$ acquired along the $[11\bar{2}0]$ zone axis (Fig. 3K-M), together with the corresponding side-view structural model (Fig. 3N), reveals a periodic stacking of atomically thin layers separated by a well-defined interlayer spacing. The projected structural model overlaid on the micrograph matches the stacking sequence expected for the $P\bar{3}m1$ structural model, in which $Tm_2C$ slabs are symmetrically terminated by F atoms and separated by weakly interacting interlayer regions with a van der Waals gap of ~1.43 Å between opposing F-terminated surfaces.

The side-view STEM images of $Pr_2CBr_2$ along the $[1\bar{1}20]$ direction (Fig. 3O–P) reveal the same layered stacking with visible Br termination contrast. The observed atomic arrangement is consistent with the structural model presented in Fig. 3Q. For $Ho_2CBr_2$, the $[11\bar{2}0]$ projection reveals pronounced contrast arising not only from the Ho planes but also from the Br termination layers (Fig. 3R-T). Because of the larger atomic number of Br relative to F, the surface terminations become visible in the STEM contrast, in agreement with the structural model shown in Fig. 3U.

In general, Br-terminated systems exhibit significantly larger separations between adjacent $M_2C$ layers than their F-terminated counterparts due to the larger size of Br and an expansion of the interlayer gap. Comparable lattice parameters are observed among the F-terminated systems, such as $Tm_2CF_2$ ($a = b = 3.61$ Å and $c = 6.25$ Å) and $Ho_2CF_2$ ($a = 3.64$ Å, $c = 6.32$ Å) (Fig. S8, Table S8), whereas $Ho_2CBr_2$ shows expanded values ($a = 3.72$ Å and $c = 9.76$ Å) (Fig. S9, Table S9). This pronounced increase of ~3.4 Å in the $c$ parameter is also evident in the $[11\bar{2}0]$ STEM projection (Fig. 3T-U). The van der Waals gap, illustrated schematically in Fig. 3T, and 3U, is also substantially larger for $Ho_2CBr_2$ (3.22 Å) than for $Ho_2CF_2$ (1.44 Å), consistent with the general trend observed for Br-terminated phases. This increased spacing indicates weaker interlayer coupling compared with F-terminated structures, which likely facilitates intercalant ingress and promotes easier delamination into MXene flakes. Consistent with this mechanism, halide-based post-treatments that replace F terminations while simultaneously expanding the interlayer spacing with Br have been shown to tune both gallery spacing and surface chemistry, thereby improving layer separation in halogen-modified MXenes (*47*). A comprehensive summary of van der Waals gaps and bond lengths for all 45 systems is provided in Table S11.

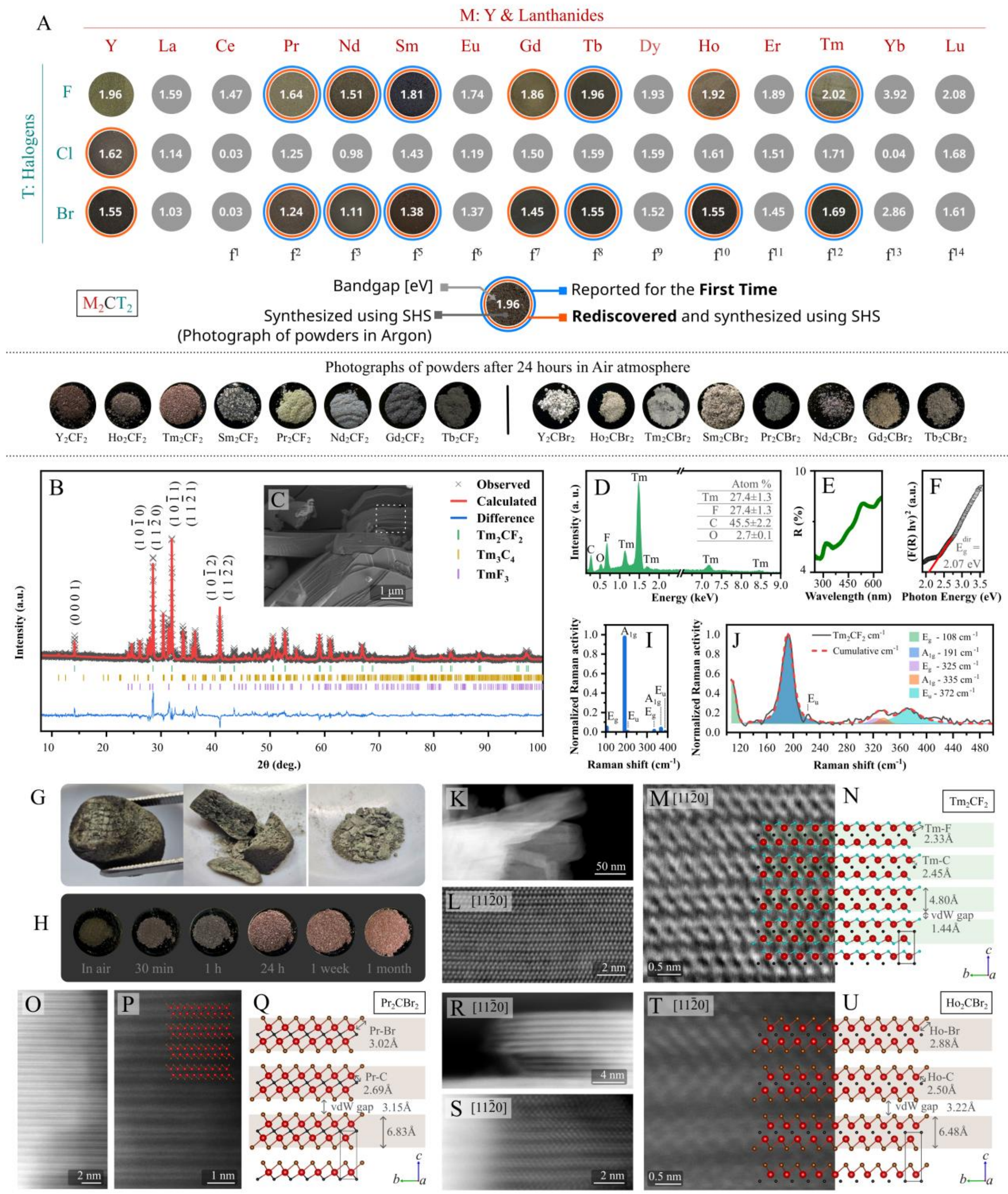

**Fig. 3** Experimental realization and characterization of rare-earth $M_2CT_2$ MXenes. (A) Overview of synthesized $M_2CT_2$ compositions highlighting SHS-synthesized structures and newly reported systems; inset values denote the theoretical band gap. Photographs of representative SHS-synthesized samples after 24 h of air exposure are shown. (B) XRD Rietveld refinement of $Tm_2CF_2$. (C) SEM image of $Tm_2CF_2$. (D) EDX spectrum of $Tm_2CF_2$. (E) UV-Vis diffuse reflectance spectrum of $Tm_2CF_2$. (F) Kubelka-Munk-derived Tauc plot of $Tm_2CF_2$. (G) Optical photograph of the SHS-synthesized $Tm_2CF_2$ pellet before and after crushing. (H) Optical

photographs showing the evolution of $Tm_2CF_2$ powder upon air exposure for several seconds (left), 30 min, 1 h, 1 day, and 1 week. (I) Calculated Raman spectrum of $Tm_2CF_2$. (J) Experimental Raman spectrum of $Tm_2CF_2$. (K-M) STEM images of $Tm_2CF_2$ acquired along the $[11\bar{2}0]$ zone axis (side view). (N) Corresponding side-view structural model derived from Rietveld refinement, highlighting the van der Waals gap, interlayer spacing, slab thickness, and selected bond lengths. (O-P) STEM images of $Pr_2CBr_2$ acquired along the $[11\bar{2}0]$ zone axis (side view). (Q) Corresponding side-view structural model. (R-T) STEM images of $Ho_2CBr_2$ acquired along the $[11\bar{2}0]$ zone axis (side view). (U) Corresponding side-view structural model derived from Rietveld refinement.

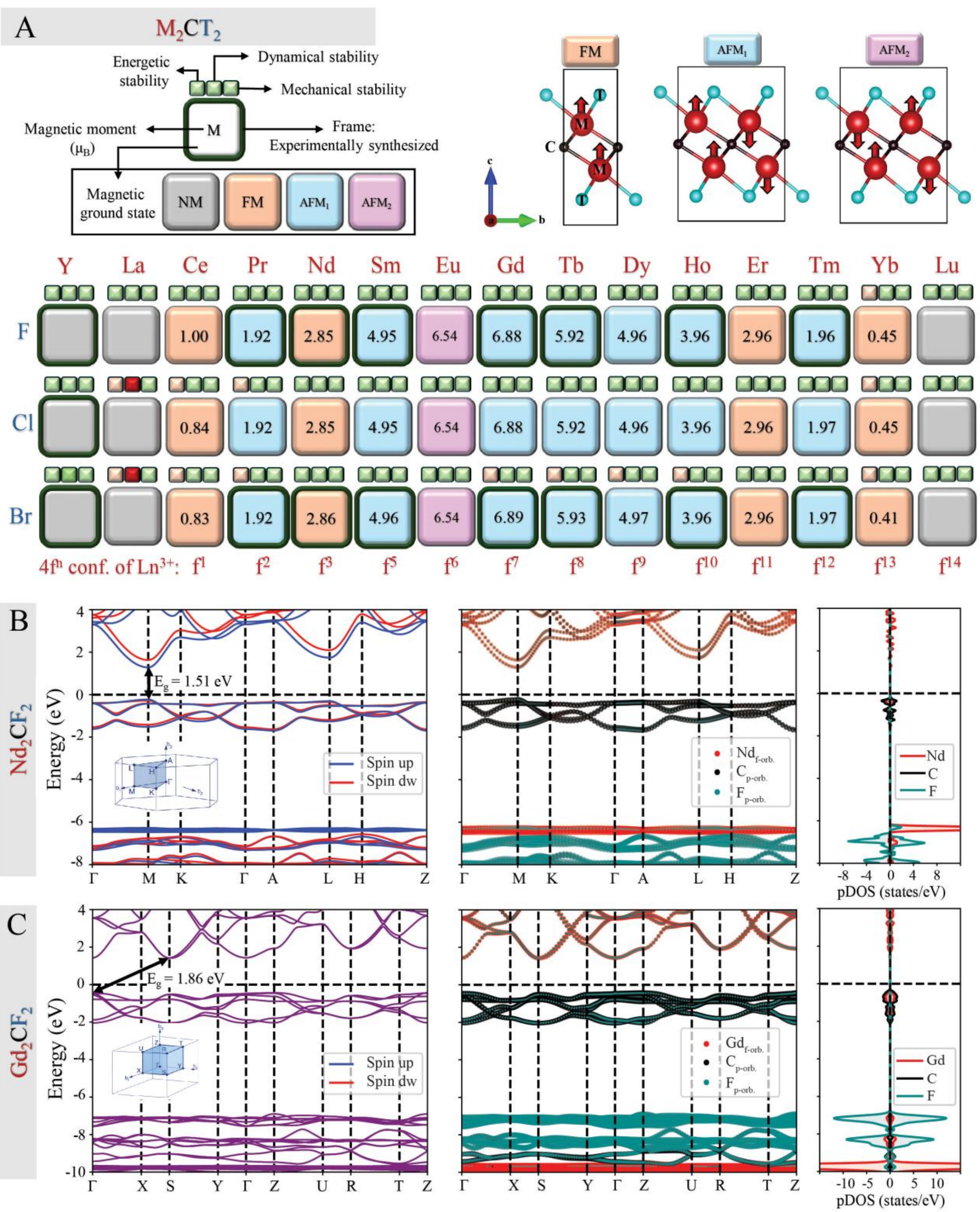

**Fig. 4.** Theoretical stability, magnetism, and electronic structure of $M_2CT_2$ MXenes. (A) Summary of theoretical results for $M_2CT_2$ compounds. Each box represents one compound, with columns and rows corresponding to the M and T elements, respectively. The three small squares above each box indicate stability criteria: the left square shows the energy above the convex hull (green = on the hull; orange = up to 0.2 eV per atom above the hull), the middle square indicates dynamical stability (green = stable; red = unstable), and the right square indicates mechanical stability (green

= stable). The color of the main box denotes the magnetic ground state (grey = non-magnetic, orange = ferromagnetic, blue/pink = antiferromagnetic). The number inside each box gives the spin magnetic moment on the M atoms in $\mu_B$. Compounds highlighted with a dark green frame have been experimentally synthesized. (B,C) Electronic band structures of two representative compounds, $Nd_2CF_2$ (FM) and $Gd_2CF_2$ (AFM), together with atom- and orbital-resolved band structures (fat bands) and projected density of states calculated using the HSE06 functional combined with DFT+U. Double-headed arrows indicate the band gap. Insets in the left panels show the high-symmetry path in the reciprocal lattice. The horizontal dashed black lines represent the Fermi level, taken as the zero-energy reference.

### *Stability and properties from first principles*

Using density functional theory (DFT), we systematically investigated 45 $M_2CT_2$ MXenes (M = Y and lanthanides; T = F, Cl, and Br) to evaluate their synthesizability and physical properties. Good agreement was observed between the theoretically optimized structures and the corresponding experimentally realized compounds (Table S11). 43 $M_2CT_2$ ml-MXenes are predicted to be stable, lying on or very close to the convex hull (within 0.2 eV/atom) while fulfilling both dynamical and mechanical stability requirements (Fig. 4A). According to the Materials Project database (*48*) many compounds within this energy range have been successfully synthesized, indicating that such energetic deviations do not necessarily preclude experimental realization (*49*). Consistent with this observation, our calculations predict energies above the convex hull of 0.12, 0.08, 0.09, and 0.07 eV per atom for $Pr_2CBr_2$, $Gd_2CBr_2$, $Tb_2CBr_2$, and $Ho_2CBr_2$, respectively, all synthesized in the present study. Imaginary vibrational frequencies were identified only for the La-based compounds with Cl and Br terminations. The computed energies above the convex hull and elastic constants are summarized in Table S12 and Fig. S34-47, respectively, while phonon spectra for each compound are provided in Fig. S23–S33. Among the 43 predicted stable compounds, 17 were here synthesized through SHS, while the remaining 26 are predicted to be promising candidates for future experimental realization.

Characteristic fingerprints, including X-ray diffraction (XRD), infrared (IR), and Raman spectra, were simulated to facilitate the identification of the target compounds. The calculated XRD, IR and Raman spectra are shown in Fig. S48–S92, while the corresponding numerical data and simulated XRD patterns are provided in the supplementary data. All investigated systems belong to the $D_{3d}$ point group, where Raman- and IR-active vibrational modes are determined by point-group symmetry (Table S13). Across the lanthanide series, the IR-active modes generally shift toward higher frequencies as the lattice contracts with increasing atomic number. The $A_{2u}$ modes (z-axis eigenvectors) are less sensitive to in-plane lattice variations than the $E_u$ modes ($x$- and $y$-components). In contrast, along the halogen series both the interlayer spacing ($c$) and the in-plane lattice parameter ($a$) increases, causing all vibrational modes to shift toward lower frequencies with stronger shift for $A_{2u}$ modes. The $E_u$ modes generally display higher intensities than the $A_{2u}$ modes in the IR spectra. A similar trend is observed for Raman-active modes: the $E_g$ modes involve in-plane atomic displacements, whereas the $A_{1g}$ mode corresponds to out-of-plane motion. In contrast to the IR case, however, the $A_{1g}$ mode exhibits the highest Raman intensity across all compounds. These results indicate that vibrational modes with dominant $z$-components are more Raman active, whereas in-plane displacements tend to produce stronger IR responses.

Rare-earth elements host large magnetic moments and are critical for a range of technologies, including permanent magnets. The magnetic ground states of the ml-MXenes were determined by evaluating the energy differences between several magnetic spin configurations, as illustrated in Fig. 4A and Fig. S93. The resulting color-coded magnetic states depend primarily on the M element, remaining consistent along columns (different surface terminations). Compounds without unpaired *f*-electrons (Y-, La-, and Lu-based) are predicted to be non-magnetic. Several compounds with an odd number of *f*-electrons (Ce-, Nd-, Er-, and Yb-based) exhibit ferromagnetic (FM) ground states, whereas the majority of the phases stabilize in antiferromagnetic (AFM) configurations. AFM ordering has been experimentally confirmed for $Ho_2CF_2$, with a Néel temperature of 3.7 K (*28*). Both FM and AFM materials play important roles in applications such as magnetic junctions, magnetic memories, and spintronic devices (*50*). Despite numerous theoretical predictions and extensive experimental efforts, intrinsic magnetism in non-alloyed transition-metal MXenes remains largely absent. Attainable AFM and FM states in RE-based $M_2CT_2$ MXenes therefore gives this materials family particular significance.

The numerical values shown in each box in Fig. 4A represent the calculated spin magnetic moments on the M atoms (units of $\mu_B$), which correlate closely with the number of unpaired *f*-electrons. A maximum moment of 6.9 $\mu_B$ is obtained for Gd-based MXenes (seven unpaired electrons). Rare-earth elements generally adopt a 3+ oxidation state, which is largely reflected in the calculated magnetic moments. Some deviations from this trend occur, most notably for Eu-based compounds where a ~6.5 $\mu_B$ moment suggests an intermediate valence resulting from partial *f*-electron delocalization and hybridization between 3+ and 2+ oxidation states (six and seven unpaired electrons). In contrast to other AFM systems, Eu-based compounds induce magnetic moments on the C atoms (> 0.2 $\mu_B$ in $Eu_2CF_2$), likely stabilizing the AFM2 configuration. Among the FM systems, Yb-based compounds exhibit similar C-induced magnetic moments (~0.4 $\mu_B$). This behavior originates from hybridization between Yb *f*-states and C *p*-states located just below the Fermi level ($E_f$) (Fig. S133).

Fig. 4B and 4C illustrate the electronic band structures of two representative compounds, $Nd_2CF_2$ (FM) and $Gd_2CF_2$ (AFM), selected from the set of synthesized materials. Electronic band structures for all synthesized and predicted compounds are provided in Fig. S94–S138 in the Supplementary Materials. A wide range of direct and indirect electronic band gaps is observed across the $M_2CT_2$ MXene family, for most systems between 1 and 2 eV, a range well-suited for integration into transistors and photovoltaic devices (*51*). Several FM systems, including $Ce_2CT_2$ and $Yb_2CT_2$ exhibit half-metallic behavior, characterized by a band gap in only one spin channel, thereby enabling potential applications as spin filters (*52*). The left panels in Fig. 4B and 4C show the spin-polarized electronic band structures for $Nd_2CF_2$ (FM) and $Gd_2CF_2$ (AFM), with insets indicating the high-symmetry paths used for the band-structure calculations. $Nd_2CF_2$ exhibits a hexagonal Brillouin zone (BZ). A pronounced spin splitting is observed in the localized *f*-states below $E_f$, a feature shared with other FM Nd-based MXenes. For $Gd_2CF_2$, the AFM ordering reduces symmetry to orthorhombic. Although $Gd_2CF_2$ displays an indirect band gap of 1.86 eV, the direct gap at the high-symmetry point "S" is very close in energy. Similarly, the flat valence band just below the Fermi level in all indirect-gap systems (Fig. S94-S138) leads to only a marginal energy difference between direct and indirect gaps, effectively rendering them semi-direct and mitigating limitations associated with indirect-gap materials in device applications. In Fig. 4 B-C, the atom- and orbital-resolved electronic states are illustrated by the fat-band representations (middle panels) and the projected densities of states, pDOS (right panels). The valence bands below $E_f$ are dominated by carbon *p*-states with moderate M hybridization, whereas the conduction bands are mainly from M states. The localized *f*-states lie deep below $E_f$ (around −6 eV for $Nd_2CF_2$ and −10 eV for $Gd_2CF_2$). Similar electronic features are observed across the $M_2CT_2$ MXene family. In half-metallic systems, localized f-states reside near the Fermi level and exhibit strong hybridization with C states, giving rise to this behavior. The general diversity in electronic structure combined with the variety of magnetic ground states, renders the $M_2CT_2$ family of MXenes attractive candidates for diverse electronic and spintronic applications.

The present work highlights the importance of expanding the MXene chemical space beyond transition metals by incorporating RE elements on the M site. This introduces localized *f*-electron states, large magnetic moments, and diverse oxidation chemistries, enabling electronic, magnetic, and optical functionalities not accessible in conventional transition-metal MXenes. The compositions compiled in the *Treasure Chest* (Table 1) further reveal a surprisingly rich multilayer chemistry, including several compounds with S or P as the X element, chemistries largely unexplored in the MXene context. Previous efforts to realize S- or P-based MXenes, primarily within Ti-based systems, have typically resulted in these elements appearing as surface terminations rather than as constituents of the MXene core, with sparce evidence of possible C substitution (*53*). The *Treasure Chest* therefore provides experimentally grounded inspiration for revisiting these chemistries using RE elements as stabilizing M components.

Altogether, the present work expands the MXene family by 49 members: 38 identified through data-driven screening and 11 synthesized here for the first time, thereby contributing to the development of broader design rules for MXene materials. Notably, the developed ML-assisted approach can be generalized to other materials families. In addition, the demonstration of direct synthesis through self-propagating high-temperature synthesis introduces a rapid and scalable route for MXene production. Such sustainable synthesis strategies will be essential for bridging the gap between fundamental laboratory discoveries and the deployment of MXenes in future technologies.

**Summary**

MXenes represent a rapidly growing class of two-dimensional materials with exceptional potential across a wide range of technologies. Here we show that the experimental literature already contains a substantial reservoir of overlooked MXene-like compounds. By combining machine-learning-assisted database mining with experiment, we identify 50 multilayer MXenes, including a “Treasure Chest” of previously synthesized materials that had not been recognized within the MXene framework. Guided by these insights, we experimentally rediscover five MXenes and realize 11 previously unexplored rare-earth (RE)-based $M_2CT_2$ MXenes, thereby expanding the experimentally accessible MXene family and adding the M elements Pr, Nd, Sm, Gd, Tb, Ho, and Tm. The incorporation of RE elements introduces new chemical and physical degrees of freedom, including semiconducting electronic structures and diverse predicted magnetic states. Complementary first-principles calculations further indicate that many additional RE MXenes are thermodynamically and dynamically stable, highlighting the large unexplored compositional space of this materials family and providing a basis for emerging design rules. The Treasure Chest compositions additionally point toward unexplored multilayer chemistries beyond the conventional C/N MXene systems. In parallel, we demonstrate a rapid and direct synthesis route based on self-propagating high-temperature synthesis, enabling MXene formation within minutes without sustained external heating. Together, these results expand the chemical landscape of MXenes and establish a data-driven strategy for uncovering functional materials hidden in existing datasets, while advancing scalable synthesis approaches that may help bridge the gap between fundamental MXene research and practical technologies.

## Acknowledgments

J.R. acknowledges funding from the European Union (ERC, MULTI2D, 101087713), and from the Knut and Alice Wallenberg (KAW) Foundation for a Scholar Grant (KAW 2023.0250). J.R and P.P. acknowledges the Wallenberg Initiative Materials Science for Sustainability (WISE) through KAW. I.H. acknowledges Estonian Research Council (Grant No. PRG3028). The computational analyses were enabled by resources provided by the National Academic Infrastructure for Supercomputing in Sweden (NAISS), partially funded by the Swedish Research Council (VR) through grant agreement no. 2022-06725. VR and the Swedish Foundation for Strategic Research (SSF) are acknowledged for access to ARTEMI, the Swedish National Infrastructure in Advanced Electron Microscopy (2021-00171 and RIF21-0026). M.D. and J.R. acknowledge funding from VR (2023-04833 and 2025-06131).

**Author contributions:** Conceptualization: J.R. Methodology: A.S., G.P., S.E., and J.R. Formal analysis: A.S., G.P., S.E., R.I., M.D., F.C., S.C., and R.T. Investigation: A.S., G.P., S.E., R.I., M.D., F.C., S.C., and R.T. Writing – original draft: A.S., G.P., S.E. and J.R. Writing – review and editing: All authors. Visualization: A.S., G.P., and S.E. Supervision: J.R., P.P., I.H., F.H. Project administration: J.R. Funding Acquisition: J.R.

**Competing interests:** The authors declare no competing interests.